# The transverse shift of a high order paraxial vortex-beam induced by a homogeneous anisotropic medium

T. A. Fadeyeva, A. F. Rubass, A. V. Volyar

*Physics Department, Taurida National V.I. Vernadsky University*
*Vernadsky av.4, Simferopol, Ukraine, 95007*
*e-mail: volyar@crimea.edu*

We consider the propagation of a tilted high order paraxial vortex-beam through a homogeneous anisotropic medium of a uniaxial crystal. We found that the initially circularly polarized beam bearing the *l*-order optical vortex splits into ordinary and extraordinary beams with a complex vortex structure. After a series of dislocation reactions the vortices gather together at the axis of the partial beam with the initial circular polarization shaping the *l*-order optical vortex. However, only *l-1* vortices gather together on the axis of the partial beam with the orthogonal circular polarization. One optical vortex is shifted along the direction perpendicular to the inclination plane of the beam. Such a vortex displacement induces the transverse shift of the partial beam. In fact, we deal with *the beam quadrefringence* in a uniaxial, homogeneous anisotropic medium. The first two beams is a result of the splitting of the initial tilted beam into the ordinary and extraordinary ones. The appearance of the second two beams is caused by the transverse shift of one of the circularly polarized components in the initially circularly polarized vortex-beam or both circularly polarized components in the initially linearly polarized beam .We consider this effect both in terms of the conservation law of the angular momentum flux along the crystal optical axis and on the base of the solutions to the paraxial wave equation for the initially circularly and linearly polarized beams. We revealed that the transverse shift of the crystal traveling beam depends on neither a magnitude nor a sign of the vortex topological charge being defined only by a handedness of the initial circular polarization and a sign of the inclination angle of the beam. Also we analyze the deformation of the cross-section of the shifted vortex-beams and its evolution as the beam propagates along the crystal.

## I. Introduction

The concept of the propagation of a plane wave (or a light ray associated with it) through a homogeneous anisotropic medium (or an unbounded crystal) presented by the Fresnel formulas [1] is a commonplace in physical optics. Transmitting obliquely to the crystal optical axis a light ray splits into two ones – the ordinary and extraordinary rays with orthogonal linear polarizations. The light velocities of these rays (or the plane waves) and directions of the electric fields are derived from the Fresnel equation. However, in most real cases, we deal with a light beam that represents a coherent bundle of rays (plane waves) with different light velocities and directions of the electric and wave vectors. Naturally, the light beam in the crystal can manifest new properties different from those in a separate plane wave. For example, an axially symmetric beam in a uniaxial crystal can be converted into an astigmatic beam [2, 3] whereas the conical refraction of the beams in a biaxial crystal embeds unique singular points (so called the diabolical points) in the beam wavefront [4]. Of even greater dramatic case is the propagation of Gaussian [5-7] and singular beams [8] along the optical axis of a uniaxial crystal. In this connection it should be noted that the singular beam (or the vortex-beam) represents a wave structure containing a set of optical vortices [9] i.e. the phase singularities of the wavefront where the field amplitude is zero while the phase is uncertain. The optical vortex is characterized by a topological charge *l* equal to a number of wavefront branches in the vicinity of the singular point. The propagation of the singular beam in free space or a homogeneous isotropic medium obeys a simple requirement: a total vortex topological charge does not change when propagating [10]. However this requirement gets broken in an anisotropic homogeneous medium.

Indeed, the right hand polarized vortex-beam with a topological charge $-l$ at the crystal input induces the orthogonally polarized beam with a topological charge $-l+2$ inside the crystal [5-8]. Such a beam carries over a spin and orbital angular momentum [11]. The spin angular momentum (SAM) is associated with a polarization of the beam field while the orbital angular momentum (OAM) is characterized by the beam structure, in particular, by the optical vortices imbedded in the beam wavefront. Although the polarization state and the vortex structure are transformed when propagating the beam, a total angular momentum flux along the crystal optical axis is conserved due to the spin-orbit coupling [12].

When tilting a vortex-beam relative to the crystal optical axis its structure experiences essential transformations [13, 14]. For example, we have recently shown [13] that if the right hand polarized (RHP) beam with a negatively charged optical vortex $l=-1$ at the crystal input propagates along the crystal optical axis, the positively charged optical vortex with $l=+1$ is imbedded in its orthogonally polarized component. At the same time, the beam inclination relative to the crystal optical axis results in splitting the initial beam into the ordinary and extraordinary ones so that both circularly polarized components get the same negatively charged vortices $l=-1$.

In fact, the beam loses two vortices, besides, it gets depolarized. Naturally, such transformations of the polarization and vortex structure cannot but affect the inner structure of the beam as a whole. We have treated these transformations in Ref 13 as splitting the initial vortex into four partial vortices and called it the vortex *quadrefringence*. However, the further consideration of this effect pointed to a more deep physical underlying of the problem that has much to do with the Fedorov-Imbert effect [15-20]. The basic part in these effects plays a conservation low of the angular momentum flux. The transformation of polarization states and the propagation direction disturbs the balance between the spin and orbital angular momenta. However, the transverse shift of the beams at the boundary face recovers this balance so that the component of the total angular momentum flux normal to the interface is conserved.

The transverse shift manifest itself also in the spin Hall effect: the splitting of a linearly polarized beam into two circularly polarized ones [21-23] and in the optical Magnus effect: rotation of the trajectory of a circularly polarized ray in an optical fiber [24, 25]. The singular beams bearing optical vortices enhance noticeably the effect owning to an additional orbital angular momentum associated with the optical vortices [26-28]. Moreover, the direction of the beam shift at the boundary face is defined now not only by the handedness of the circular polarization but also by the sign of the vortex topological charge

In the case of the vortex-beam separation inside a homogeneous anisotropic medium, variations of the propagation direction and the beam depolarization transform the beam state [13] in a similar manner as that happens in the mentioned above Fedorov-Imbert effect. However, the major difference lies in the fact that the vortex quadrefringence occurs inside a homogeneous anisotropic medium and is accompanied by the transformation of the vortex structure while the classical transverse shift needs a non-homogeneous medium or an interface between two homogeneous media, the vortex structure does not changing essentially.

In the present paper we consider a new manifestation of the transverse shift of the vortex-beams induced by a homogeneous anisotropic medium; in particular, we will focus a special attention on the destruction and recovery of the singular beams bearing high order optical vortices.

## II. Preliminary consideration: spin-orbit coupling in a homogeneous anisotropic medium

We will consider an unbounded homogeneous anisotropic medium with the only optical axis directed along the z-axis of the referent frame x,y,z (Fig.1). A paraxial vortex-beam with the electric field $\mathbf{E}(x,y,z)$ propagates at some small angle $\alpha_o$ relative to the crystal optical axis. We assume the refractive index of the ordinary beam $n_o$ to be larger than that of the extraordinary one $n_e$: $n_o > n_e$. Let the beam has the only RHP component $E_+(x,y,z=0)$ at the plane z=0 and carries over a negatively charged optical vortex (l=-1) positioned at the beam axis while the amplitude of the left hand polarized (LHP) beam component vanishes: $E_-(x,y,z=0,)=0$. When transmitting the circularly polarized beam $E_+$ along the crystal optical axis, the anisotropic medium induces the second vortex-beam $E_-(x,y,z)$ in the orthogonally polarized component propagating along the same direction, its topological charge being equal to l=+1 [5,7]. The inclination of the initial beam $\mathbf{E}(x,y,z)$ on the y0z plane at a small angle $\alpha_o \ll 1$ results in splitting each circularly polarized beam component into two partial beams whose axes are tilted at the different angles $\alpha_o$ and $\alpha_e$, respectively. After a series of topological reactions, all partial beams acquire the optical vortices with the same topological charges l=-1 [13, 14]. At the first glance it seems that each pair of beams with orthogonal circular polarizations propagating at the same angle $\alpha_o$ (or $\alpha_e$) forms the vortex-beams with uniformly distributed linear polarization at the beam cross-section. However, the positions of optical vortices in these components do not coincide with each other at any crystal length and the field gets non-uniformly polarized in the vicinity of singular points. The vortices are always shifted relative to each other at the distance $\Delta x_V = -2/(k_o \alpha_o)$ along the direction perpendicular to the plane of the beam inclination. We called this effect the vortex *quadrefringence* [13]. Let us consider this process in terms of conservation law of the angular momentum flux.

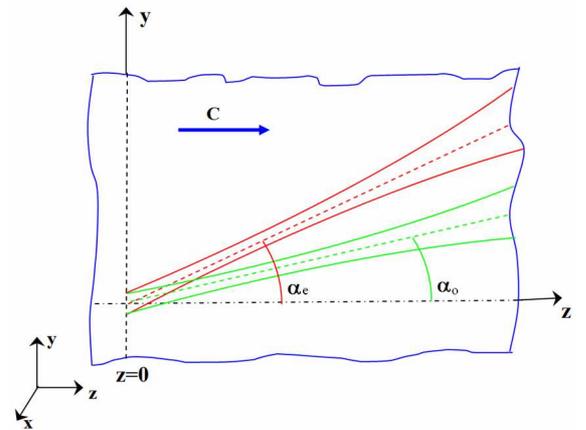

*Fig.1 (online) The sketch of the beam propagation in a uniaxial crystal. **C** is a unit vector of the crystal optical axis*

Generally speaking, the conservation law of the angular momentum of light in a simple form for homogeneous isotropic media [11] cannot be employed for an anisotropic medium because the anisotropic crystal has sources and sinks of the angular momentum of light [29] that is the angular momentum of light is coupled with the angular momentum of the medium. Nevertheless, Ciattoni et al. [12] showed that the conservation law can be written for the component of the total angular momentum flux along the optical axis of the uniaxial crystal where the medium is rotationally invariant and the coupling between the angular momentum of the medium and light vanishes. The balance equation can be written in the form:

$$S_z + L_z = I_+ - I_- + L_z^{(+)} + L_z^{(-)} = const, \qquad (1)$$

where $S_z$ and $L_z$ stand for the spin and orbital angular momenta, respectively, $I_+$ and $I_-$ are the dimensionless intensities of the RHP and LHP components, $L_z^{(+)}$ and $L_z^{(-)}$ are the OAM of the RHP and LHP components. Although the spin and orbital components of the angular momentum flux can change their magnitudes when transmitting the beam, their sum remains constant. Indeed, if the beam has a uniformly distributed circular polarization at the initial plane z=0, its polarization degree decreases gradually when propagating the beam along the crystal optical axis so that ultimately the beam is totally depolarized at a large crystal length [12]. Its SAM vanishes. However, a spin-orbit coupling stimulates a gradual growth of the OAM, its magnitude culminating in the asymptotic case. The beam inclination entails dislocation reactions in each circularly polarized component [13]. Their intensities $I_+$ and $I_-$ begin to oscillate very quickly drawing the orbital and spin components of the angular momentum into these fast oscillations. The events of the birth and annihilation of optical vortices in the beam make the center of gravity of each beam component to vibrate. Besides, two vortices get lost [13] leaving the LHP component. Naturally the vortex losses also contribute to the OAM. As the light wave propagates along the crystal, the partial beams move away from each other and the oscillations die down. However, the centers of gravity of the individual beams do not revert to the common plane formed by the initial beam axis and the crystal optical axis. The beam axes have to be shifted relative to this plane because of the loss of two optical vortices and the light depolarization.

Let us estimate the magnitude of the lateral shift $\Delta x_T$ of the above considered tilted vortex-beam (see Fig.2) on the base of the balance equation (1). We rewrite at first the equation (1) in terms of the center of gravity of the beam [17]:

$$(\mathbf{r}_c \times \mathbf{k}_c)_z + (I_+ - I_- + I_+ l_+ + I_- l_-)\frac{k_z}{k} = const, \quad (2)$$

where $\mathbf{r}_c$ is the radius vector of the center of gravity, $k_z$ stands for the z-component of the wave vector of the beam associated with the center of gravity, $l_+$ and $l_-$ are the vortex topological charges in the RHP and LHP components, respectively. Since the inclination angle is small we can assume that $k_z/k \approx 1$. At the z=0 plane, terms of eq. (2) for our vortex-beam can be defined as: $I_+(z=0)=1$, $I_-(z=0)=0$, $(\mathbf{r}_c \times \mathbf{k}_c)_z = 0$, $l_+ = -1$. In the asymptotic case for the crystal length $z \to \infty$, the light beam is depolarized: $I_+(z) = I_-(z) = 0.5$. Besides, the partial beams in the RHP and LHP component acquire the negatively charged vortices: $l_+ = l_- = -1$ and, besides, $(\mathbf{r}_c \times \mathbf{k}_c)_z = -k_o \alpha_o \Delta x_T$ in the referent frame shown in Fig.1. Equating the momentum fluxes at the z=0 plane and the $z=const$ plane ($z \to \infty$) one comes to the equation $1 - 1 = 0.5 - 0.5 - 0.5 - 05 - k_o \alpha_o \Delta x_T$ or

$$\Delta x_T = -\frac{1}{k_o \alpha_o} = -\frac{\lambda}{2\pi n_o \alpha_o}, \quad (3)$$

Since the vortex structure in the $E_+$ component does not change in the asymptotic case, we can assume its transverse shift to be zero: $\Delta x_+ = 0$ while the lateral shift of the left hand component $\Delta x_-$ to be twice as much as the total shift $\Delta x_T$: $\Delta x_- = 2\Delta x_T$. The equation (3) indicates that the lateral shift is proportional to the wavelength $\lambda$ and inversely proportional to the inclination angle $\alpha_o$ of the center of gravity. It might be well to point out that the eq. (3) was obtained under the condition that the partial beams in the field components were separated, i.e. the angle $\alpha_o$ cannot be zero: $\alpha_o \neq 0$.

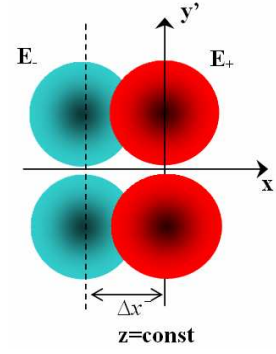

*Fig.2 (online) The transverse shift $\Delta x^-$ of the LHP $E^-$ beam*

In the equation (2), we cannot change arbitrary the magnitudes and signs of the vortex topological charges in the right and left hand components of the initial beam in the crystal (in contrast to those in free space) because of their complex structure (see bellow). As a result the equation (2) in the above example does not take into account the influence of these beam characteristics on the transverse shift $\Delta x_T$. Besides, this equation does not permit us to find the transverse shift of each circularly polarized component: $\Delta x^+$ and $\Delta x^-$. To describe such a process in details it is necessary to analyze the solutions to the paraxial wave equation.

### III. Destruction and recovery of high order vortex beams

#### III.1 The basic groups of vortex-beams

The anisotropic uniaxial medium is described by the permittivity tensor in the form:

$$\hat{\varepsilon} = \begin{pmatrix} \varepsilon & 0 & 0 \\ 0 & \varepsilon & 0 \\ 0 & 0 & \varepsilon_3 \end{pmatrix}. \quad (4)$$

We assume that the paraxial beam propagates along the crystal optical axis: $\mathbf{E}(x,y,z) = \tilde{\mathbf{E}}(x,y,z)\exp(-i k_o z)$, where $k_o = n_o k_0$ is a wavenumber of the ordinary beam in the crystal, $n_o^2 = \varepsilon$, $k_0$ is a wavenumber in free space. The paraxial wave equation for the transverse component of the electric vector $\tilde{\mathbf{E}}_\perp = \mathbf{e}_x \tilde{E}_x + \mathbf{e}_y \tilde{E}_y$ ($\mathbf{e}_x, \mathbf{e}_y$ are the unit vectors) can be written in the form [2, 8,13]:

$$(\nabla_\perp^2 - 2i k_o \partial_z)\tilde{\mathbf{E}}_\perp = \beta \nabla_\perp (\nabla_\perp \tilde{\mathbf{E}}_\perp), \quad (5)$$

where $\beta = \Delta\varepsilon/\varepsilon_3$, $\Delta\varepsilon = \varepsilon_3 - \varepsilon$. Let us make use of new coordinates: $u = x + i y$, $v = x - i y$ and a new polarization basis: $\tilde{E}_+ = \tilde{E}_x - i\tilde{E}_y$, $\tilde{E}_- = \tilde{E}_x + i\tilde{E}_y$. Then the eq. (5) can be rewritten as

$$(4\partial_{uv}^2 - 2i k_o \partial_z)\tilde{E}_+ = 2\beta \partial_u (\partial_v \tilde{E}_+ + \partial_u \tilde{E}_-), \quad (6a)$$

$$(4\partial_{uv}^2 - 2i k_o \partial_z)\tilde{E}_- = 2\beta \partial_v (\partial_v E_+ + \partial_u E_-). \quad (6b)$$

Particular solutions to the eqs (6) can be found by means of simple substitutions:
1) the ordinary mode beam:

$$\tilde{E}_+^o = w_0 \partial_u \Psi_o, \quad \tilde{E}_-^o = -w_0 \partial_v \Psi_o, \quad (7)$$

2) the extraordinary mode beam:

$$\tilde{E}_+^e = w_0 \partial_u \Psi_e, \quad \tilde{E}_-^e = w_0 \partial_v \Psi_e \quad (8)$$

The scalar function $\Psi$ is a solution to a scalar paraxial wave equation:
$$\left(\nabla_\perp^2 - 2ik_{o,e}\partial_z\right)\Psi_{o,e} = 0 \qquad (9)$$
where $k_e = \left(n_3^2/n_o\right)k_0$, $n_3^2 = \varepsilon_3$, $w_0$ is the beam waist at the plane $z = 0$.

A small inclination of the ordinary beam axis relative to the crystal optical axis (say, on the y0z plane) at a small angle $\alpha_o \ll 1$ can be taken into account by a shift of the origin of the coordinates along the imaginary y-axis at the distance $y_o = i\alpha_o z_o$ [13, 30, 31], where $z_o = k_o w_0^2/2$. The new transverse coordinates are $\bar{x} = x$, $\bar{y} = y + i\alpha_o z_o$, $\bar{u} = u - \alpha_o z_o$, $\bar{v} = v + \alpha_o z_o$. In this case, the paraxial ordinary beam with a Gaussian envelope tilted relative to the z-axis at the angle $\alpha_o$ is transformed into the beam propagating along the z-axis but its intensity maximum is shifted at the distance $y'_o = y - \alpha_o z$. The extraordinary beam is tilted at the angle $\alpha_e$ and its y-coordinate is shifted at the distance $y_e = i\alpha_e z_e$ ($z_e = k_e w_0^2/2$). The transformations of the phase in the tilted extraordinary beam are taken into account by the additional curvatures of the wavefront stimulated by the complex shift $i\alpha_e z_e$. Besides, the function of the Gaussian envelopes get an amplitude factors: $\exp(-k_{o,e} z_{o,e} \alpha_{o,e}^2/2)$. Clearly the solutions (7, 8) to the equations (9) in the coordinates $x, y, z$ also satisfy to these equations in the coordinates $\bar{x}, \bar{y}_{o,e}, z$.

It is worth to note that the operator $\partial_{\bar{u}} = \partial_u = \partial_x - i\partial_y$ acting on the generatrix function of the beam $\Psi_0$ in free space [34] is the operator of the birth of the vortex-beams with a negative topological charge $l < 0$ while the operator $\partial_{\bar{v}} = \partial_v = \partial_x + i\partial_y$ is the operator of the birth of the positively charged vortex-beams $(l > 0)$. Thus, by acting the operators $\partial_u$ and $\partial_v$ on a generatrix functions $\Psi_0^{(o,e)}$ in the crystal we can create the first large group of the high order vortex-beams. As a generatrix function $\Psi_0^{(o,e)}$ we can choose the functions of the fundamental Gaussian beam in a homogeneous isotropic medium with refractive indices $n_o$ and $n_e$, respectively:
$$\Psi_0^{(o,e)} = \frac{1}{\sigma_{o,e}}\exp\left[-\frac{\bar{u}\bar{v}}{w_0^2\sigma_{o,e}} - k_{o,e}z_{o,e}\frac{\alpha_{o,e}^2}{2}\right], \qquad (10)$$
where $\sigma_{o,e} = 1 - iz/z_{o,e}$. Thus the fields of the high order vortex-beams of the first group **V** are
$$V_+^{(-l),(o,e)} = N_l^V \partial_u^{l-1}\partial_u \Psi_0^{(o,e)},$$
$$V_-^{(-l),(o,e)} = \mp N_l^V \partial_u^{l-1}\partial_v \Psi_0^{(o,e)}, \qquad (11)$$
where $N_l^V = (-w_0)^l$, $l \geq 1$ stands for the modulus of the vortex topological charge in the RHP beam component. The sign $(-)$ corresponds to the ordinary beam $\mathbf{V}^{(-l),(o)}$. All these beams carry over optical vortices. However, this group of vortex-beams *does not involve a fundamental Gaussian beam in one of the circularly polarized components*. To create such a mode beam let us make use of eqs (7, 8) and write
$$G_+^{(o,e)} = -\int_{\bar{u}}^{\infty}\partial_u \Psi_0^{(o,e)} du \qquad (12)$$
$$G_-^{(o,e)} = \pm\int_{\bar{u}}^{\infty}\partial_v \Psi_0^{(o,e)} du \cdot \qquad (13)$$
where the sign $(+)$ refers to the ordinary beam $G_-^{(o)}$ while the sign $(-)$ is associated with the extraordinary beam $G_-^{(e)}$. The $G_-^{(o,e)}$ component in eq.(12) has an amplitude singularity at the axis: $u = v = 0$ [8]. To avoid the amplitude uncertainty we form the field as
$$U_+^{(0)} = G_+^{(o)} + G_+^{(e)}, \quad U_-^{(0)} = G_-^{(o)} + G_-^{(e)}. \qquad (14)$$
The equations (12)-(14) enable us to build the second large U-group of the vortex-beams as:
$$U_+^{(l)} = N_l^U \partial_v^l\left(G_+^{(o)} + G_+^{(e)}\right),$$
$$U_-^{(l)} = N_l^U \partial_v^l\left(G_-^{(o)} + G_-^{(e)}\right). \qquad (15)$$
where $N_l^U = (-w_0)^l$. For example, the lowest order beam with a complex amplitudes:
$$U_+^{(0)} = \Psi_0^{(o)} + \Psi_0^{(e)},$$
$$U_-^{(0)} = -\frac{\bar{u}}{\bar{v}}\left[w_0^2\frac{\sigma_o \Psi_0^{(o)} - \sigma_e \Psi_0^{(e)}}{\bar{u}\,\bar{v}} + \Psi_0^{(o)} - \Psi_0^{(e)}\right] \qquad (16)$$
carries over a centered double-charged optical vortex with $l = +2$ in the LHP component when propagating along the z-axis ($\alpha_o = 0$). Similar to the V-group of the singular beams, the second U-group carries over the optical vortices in each circularly polarized component (excepting the $U_+^{(0)}$ component) whose topological charges differ to two units. At the same time, the RHP components of the beams of the V-group carry over the negatively charged vortices while the vortices of the beams of the U-group have positive topological charges.

However, the major distinctive feature of the fields of the V-group is their transverse structure: the electric or magnetic z-components of the field are zero. Indeed, the z-component of the electric field of the ordinary beam $\mathbf{V}^{(-l),(o)}$ is defined as
$$E_{z,o}^{(-l)} \approx \frac{i}{k_o}\frac{\varepsilon e^{-ik_o z}}{\varepsilon_3}\nabla_\perp \mathbf{V}^{(-l),(o)} =$$
$$= \frac{i}{k_o}\frac{\varepsilon e^{-ik_o z}}{\varepsilon_3}N_l^V \partial_u^{l-1}\left(\partial_{uv}^2\Psi_0^{(o)} - \partial_{vu}^2\Psi_0^{(o)}\right) = 0 \qquad (17)$$
i.e. we deal with the transverse electric mode beam. The z-component of the magnetic field of the extraordinary beam is
$$H_{z,e}^{(-l)} = \frac{ie^{-ik_e z}}{k_0}\left(\nabla \times \mathbf{V}^{(-l),(e)}\right)_z =$$
$$= \frac{ie^{-ik_e z}}{k_0}N_l^V \partial_u^{l-1}\left(\partial_{vu}^2\Psi_0^{(e)} - \partial_{uv}^2\Psi_0^{(e)}\right) = 0 \qquad (18)$$
i.e. we deal with the transverse magnetic mode beam. On the other hand, the fields of the U-group have a hybrid structure and cannot be divided into ordinary and extraordinary beam fields when propagating along the crystal optical axis.

More complex singular beams $\mathbf{V}^{(-l,m)}$ and $\mathbf{U}^{(l,m)}$ with the second radial index $m$ can be derived from eqs (11) and (15) by means of the action of the operator $\hat{M}_m^{(o,e)} = (-i z_{o,e})^m m! \partial_z^m$ on the initial vectors $\mathbf{V}^{(-l)}$ and $\mathbf{U}^{(l)}$ [8].

Our major requirement to the fields $\mathbf{V}$ and $\mathbf{U}$ *is that their left hand components at the plane z=0 vanish*: $V_-^{(-l)}(x,y,z=0) = 0$, $U_-^{(l)}(x,y,z=0) = 0$. It means that the vector functions $\mathbf{V}^{(-l,m)} = \mathbf{V}^{(-l,m),(o)} + \mathbf{V}^{(-l,m),(e)}$ together with $\mathbf{U}^{(l,m)} = \mathbf{U}^{(l,m),(o)} + \mathbf{U}^{(l,m),(e)}$ define unambiguously the crystal traveling beams. It permits also us to form the arbitrary beam field $\mathbf{E}$ at the plane $z=0$ with a complex amplitude $\mathbf{W}$ in terms of the crystal-traveling beams $\mathbf{U}^{(l,m)}$ and $\mathbf{V}^{(-l,m)}$ as:

$$\mathbf{W}(x,y,z=0) = \sum_{l,m}\left[a_{l,m}\mathbf{V}^{(-l,m)}(x,y,z=0) + b_{l,m}\mathbf{U}^{(l,m)}(x,y,z=0)\right], \quad (19)$$

where $a_{l,m}$ and $b_{l,m}$ are the expansion coefficients. In our further consideration we restrict ourselves only to the vortex-beams with a zero radial index $m=0$. Besides, the above requirement enables us to obtain the relation between the angles $\alpha_o$ and $\alpha_e$ [13]:

$$k_o \alpha_o^2 z_o = k_e \alpha_e^2 z_e,$$

### III.2 Structural transformations in the tilted vortex-beams

The most intriguing feature of the tilted beams (the beams with complex variable $\bar{y} = y + i\alpha_o z_o$) of the *V*- and *U*-groups is that the optical vortex does not follow the beam axis. For example, the ordinary vortex beams of the *V*-group derived from eqs (10, 11) have the RHP component in the form

$$V_+^{(-l),(o)} = \left[\frac{x - i(y + i\alpha_o z_o)}{w_0 \sigma_o}\right]^l \frac{\exp\left[-\frac{x^2 + \bar{y}^2}{w_0^2 \sigma_o} - k_o \frac{\alpha_o^2}{2} z_o\right]}{\sigma_o}. \quad (20)$$

The Gaussian envelope has a maximum at the point $x=0$, $y = \alpha_o z$ while the vortex is positioned at the point: $y = 0$, $x = -\alpha_o z_o$. The vortex leaves the axis when tilting the beam. In order to force the vortex to follow the beam it is necessary to construct a new beam

$$V_+'^{(-l),(o)} = \left[\frac{x - i(y - \alpha_o z)}{w_0 \sigma_o}\right]^l \frac{\exp\left[-\frac{x^2 + \bar{y}^2}{w_0^2 \sigma_o} - k_o \frac{\alpha_o^2}{2} z_o\right]}{\sigma_o} \quad (21)$$

in terms of the crystal-traveling beams. Since

$$\left[\frac{x - i(y - \alpha_o z)}{w_0 \sigma_o}\right]^l \Psi_0^{(o)} = \left[\frac{x - i(y - \alpha_o z + i\alpha_o z_o - i\alpha_o z_o)}{w_0 \sigma_o}\right]^l \Psi_0^{(o)} =$$
$$= \left[\frac{x - i(y + i\alpha_o z_o)}{w_0 \sigma_o} - \frac{\alpha_o z_o}{w_0}\right]^l \Psi_0^{(o)}$$

consequently, the equation (21) is a solution to the paraxial wave equation or otherwise

$$V_+'^{(-l),(o)} = \sum_{p=0}^l \binom{l}{p}\left(-\frac{\alpha_o z_o}{w_0}\right)^{l-p}\left[\frac{x - i(y + i\alpha_o z_o)}{w_0 \sigma_o}\right]^p \Psi_0^{(o)}. \quad (22)$$

i.e. the RHP component of the tilted beam represents a superposition of elementary beams with a complex $\bar{y}$ variable. By the similar way we can write the beam components $V_+'^{(-l),(e)}$, $U_+'^{(l),(o)}$, $U_+'^{(l),(e)}$. Then with the help of the expressions (11) and (15) we build the LHP components $V_-'^{(-l),(o)}$, $V_-'^{(-l),(e)}$, $U_-'^{(l),(o)}$, $U_-'^{(l),(e)}$ and the fields $\mathbf{V}'^{(-l)}$ and $\mathbf{U}'^{(l)}$ of the crystal-traveling beams. The new beam field $V_+'^{(-l)}$ comprises not only the field of the first *V*-group but also the beam of the second *U*-group (see the term with $p=0$ in eq. (22)) whereas the beams $U_+'^{(l)}$ are defined only by the elementary beams of the *U*-group. Thus, by using expressions (11) and (15) we can write a total rule for constructing the tilted vortex beams in a uniaxial crystal.

$$V_+^{(-l)} = \left[\frac{x - i(y - \alpha_o z)}{w_0 \sigma_o}\right]^l \Psi_0^{(o)} + \left[\frac{x - i(y - \alpha_e z)}{w_0 \sigma_e}\right]^l \Psi_0^{(e)}, \quad (l \neq 0) \quad (23)$$

$$V_-^{(-l)} = -N_l^V \sum_{p=1}^l \binom{l}{p}\left(-\frac{\alpha_o z_o}{w_0}\right)^{l-p} \partial_u^{p-1}\left(\partial_v \Psi_0^{(o)} - \partial_v \Psi_0^{(e)}\right) +; \quad (l \neq 0) \quad (24)$$

$$+ \left(-\frac{\alpha_o z_o}{w_0}\right)^l U_-^{(0)}$$

$$U_+^{(l)} = \left[\frac{x + i(y - \alpha_o z)}{w_0 \sigma_o}\right]^l \Psi_0^{(o)} + \left[\frac{x + i(y - \alpha_e z)}{w_0 \sigma_e}\right]^l \Psi_0^{(e)}, \quad (25)$$

$$U_-^{(l)} = N_l^U \sum_{p=0}^l \binom{l}{p}\left(\frac{\alpha_o z_o}{w_0}\right)^{l-p} \partial_v^p \left(G_-^{(o)} + G_-^{(e)}\right). \quad (26)$$

Without loss of generality, for our theoretical analysis, we choose the initial vortex beams of the *V*- and *U*-groups with topological charges $l_V^+ = -3$ and $l_U^+ = +3$, respectively, in the RHP component at the $z=0$ plane. Then LHP components of theses groups have the vortex charges: $l_V^- = -1$ and $l_U^- = 5$, respectively, in the immediate vicinity of the $z=0$ plane. When the inclination angle $\alpha_o$ is zero $(\alpha_o = 0)$, the $V_-^{(-l)}$ and $U_-^{(l)}$ components carry over these vortices along all the z-axis [8]. However, even a very small beam inclination $(\alpha_o \neq 0)$ destroys fairly rapidly such vortex states and a total topological charge in the left hand component goes to the charge in the right hand component. Typical patterns of the intensity distributions for different crystal lengths $z$ at the given inclination angle $\alpha_o$ for the LHP components $V_-^{(-3)}$ and $U_-^{(3)}$ are shown in Fig.3. The range of the crystal lengths from $z=0$ to $z=z_{ind}$ is characterized by a chain of dislocation reactions: a succession of the events of the vortex birth and annihilation. All vortices are intermingled in the beam. We cannot speak which partial beam (ordinary or extraordinary) the group of the vortices belongs to. In the vicinity of the plane $z=z_{ind}$ (so-called the indistinguishability limit [35]) a part of vortices leaves the area of the dislocation reactions and start to form the core of both the ordinary and extraordinary partial beams. Starting from this point we can speak about splitting each circularly polarized component into the ordinary and extraordinary partial beams. The magnitude of the indistinguishability border $z_{ind}$ (or $\alpha_{ind}$) can be analytically found only for a simple case of a singly-charged vortex-beam [35]. In the rest cases, we found this beam parameter on the base of numerical calculations. Such a vortex behavior is typical also for the $V_+^{(-3)}$ and $U_+^{(3)}$ beam components (the figures are not shown). However, the vortex evolution behind the indistinguishability border $z=z_{ind}$ is radically different for them.

Let us estimate the asymptotic behavior of the vortices in the $V_+^{(-l)}$ and $V_-^{(-l)}$ components. We will assume that the ordinary and extraordinary beams in this component are completely split and the beams do not interfere with each other, i.e. $\alpha_o z/w_0, \alpha_e z/w_0 \gg 1$. Besides, we will treat the partial beam near its axis so that $|x/Z_{o,e}|, |(y-\alpha_{o,e}z)/Z_{o,e}| \ll 1$, and $Z_{o,e} = z + i z_{o,e}$, $y'_{o,e} = y - \alpha_{o,e} z$,

$$\left[\frac{1}{x^2 + (y+i\alpha_o z_o)^2}\right]_{o,e} \approx$$

$$\approx -\frac{1}{(\alpha_{o,e} Z_{o,e})^2} \sum_{q=0}^{l} (-1)^q \left(\frac{r'^2_{o,e}}{\alpha^2_{o,e} Z^2_{o,e}} + 2i \frac{y'_{o,e}}{\alpha_{o,e} Z_{o,e}}\right)^q$$

$$\left(\frac{\bar{u}}{\bar{v}}\right)_{o,e} \approx -\left[1 - \frac{i}{\alpha_{o,e} Z_{o,e}}(x + i y'_{o,e})\right]^p \times$$

$$\times \sum_{p=0}^{l} \left(\frac{-i}{\alpha_{o,e} Z_{o,e}}\right)^p (x - i y'_{o,e})$$

where $r'^2_{o,e} = x^2 + y'^2_{o,e}$. In the above equations, we restricted ourselves to the $l$-th term in the power series associated with a topological charge $l$ of the considered vortex-beam. Besides, the value $\bar{r}^2 = x^2 + (y+i\alpha_o z_o)^2$ does not depends on the $o-$ or $e-$ indices in the complex beam because $\alpha_o z_o = \alpha_e z_e$. However, it is necessary to take into account these indices in the radius $r'_{o,e}$ when considering each partial beam separately. After a tedious but straightforward algebra in the expressions (23) and (24) we come to the asymptotic expressions for the $V_+^{(-l),(o)}$ and $V_-^{(-l),(o)}$ components:

$$V_+^{(-l),o} \propto \left(\frac{x - i y'_o}{w_0 z/z_o}\right)^l \Psi_0^{(o)}, \quad (27)$$

$$V_-^{(-l),o} \propto \left(\frac{x - i y'_o}{w_0 z/z_o}\right)^{l-1} \left(\frac{z_o}{w_0 z}\right)\left(\frac{2l}{\alpha_o k_o} + x - i y'_o\right)\Psi_0^{(o)}. \quad (28)$$

Similarly we can obtain the asymptotic expressions for the extraordinary $V_+^{(-l),(e)}$ and $V_-^{(-l),(e)}$ partial beams Thus, the $V'^{(-l)}_+$ component has two branches of the ordinary and extraordinary beams with centered $l$-order optical vortices. The $V'^{(-l)}_-$ component has also two branches. However, the vortices have a complex structure. They do not gather together at the axis like those in the ordinary beam. The vortices of the $l$-1 order are positioned at the axes of the beams: $x_1^{(o)} = 0$, $y_1^{(o)} = \alpha_o z$ and $x_1^{(e)} = 0$, $y_1^{(e)} = \alpha_e z$. The second pair of vortices with unit topological charges is shifted along the x-axis at the distance $\Delta x_V = -2l/(\alpha_o k_o)$ relative to their neighbors in the $V_+$ beam component: $x_2^{(o)} = -2l/(\alpha_o k_o)$, $y_2^{(o)} = \alpha_o z$ and $x_2^{(e)} = -2l/(\alpha_o k_o)$, $y_2^{(e)} = \alpha_e z$. The magnitude of the transversal vortex shift $\Delta x$ increase linearly with growing the vortex topological charge $l$ and does not depend on the crystal length $z$. It means that the extraordinary beam do not recover its initial structure at any crystal length. It lead to a drastic consequence.

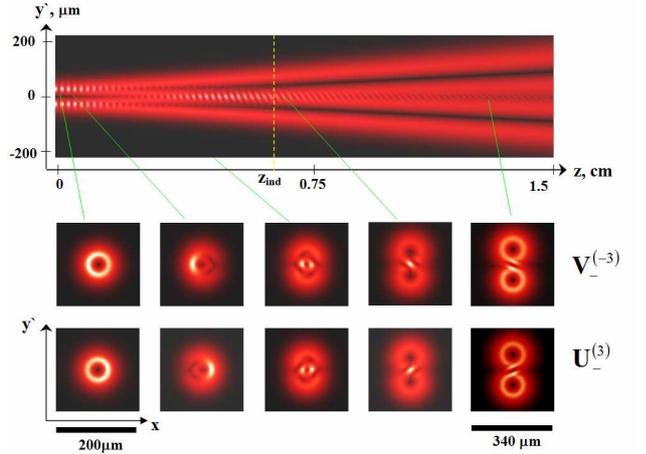

*Fig. 3 (online) Splintering of the $V_-^{(-3)}$ and $U_-^{(3)}$ beam components in the LiNbO$_3$ crystal at the inclination angle $\alpha_o = 10^o$ and $w_0 = 30\mu m$. The upper figure represents the longitudinal section of the $V_+^{(-3)}$ beam component*

Indeed, in frames of the approach of the plane waves spreading along the crystal [1] the superposition of $V_+^{(-l)}$ and $V_-^{(-l)}$ beam components must form a total beam with a uniformly distributed linear polarization over cross-sections of the splintered partial beams, the linear polarizations being orthogonal to each other in these beams. However, a complex vortex structure in each polarized component of a total paraxial wave field results in a non-uniformly polarized field distribution in the vicinity of the beam core. This situation is shown in Fig.4. The polarization distribution represents a set of polarization ellipses on the background of the polarization ellipticity $Q = \pm b/a$ ($a$ and $b$ is the ellipse axes). The solid lines (streamlines) are oriented along the major axes of the ellipses that are characterized by inclination angle $\psi$ to the x-axis [36]. The streamlines trace the characteristic pattern in the vicinity of the C-points – the points of the polarization singularity. One of the circularly polarized components vanishes at this point. In fact, the C-point characterizes the vortex position in one of the field components. There are three types of patterns traced by the streamlines: the star, the lemon and the monstar. The star is characterized by the topological index $\nu = -1/2$ whereas the lemon and the monstar have the same topological indices $\nu = +1/2$. The picture in Fig.4b has six characteristic patterns: three stars and three lemons for the one of the beams of the V-group at the inclination angle $\alpha_o = 8^o$. As the angle $\alpha_o$ increases, three lemons and two stars draw together forming the pattern with a topological index $\nu = -1/2 - 1/2 + 1/2 + 1/2 + 1/2 = +1/2$ that we classify as a degenerated lemon (see Fig. 4c). This polarization singularity corresponds to the position of the double negatively charged optical vortex in the $V_-^{(3)}$ component and the triple negatively charged vortex in the $V_+^{(-3)}$ component. The star shifted relative to the lemon corresponds to the singly charged vortex in the $V_-^{(-3)}$ component. The computer simulation showed that these polarization singularities are always separated at any crystal lengths.

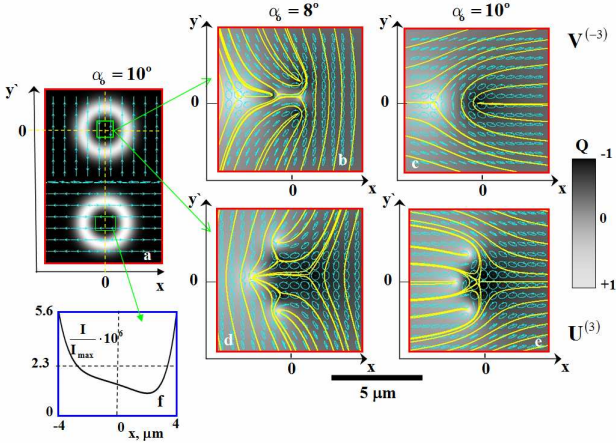

*Fig.4 (online) Polarization distribution in the splintered beams in the LiNbO$_3$ crystal with $n_1 = 2.3$, $n_3 = 2.2$, $z = 2\,cm$, $w_0 = 30\,\mu m$: (a) the intensity distribution in the vicinity of the $\mathbf{V}^{(-3)}$ beam core, (b), (c) polarization singularities for $\mathbf{V}^{(-3)}$ beam, (d), (e) polarization singularities for $\mathbf{U}^{(3)}$ beam, (f) the intensity profile in the vicinity of the polarization singularities of the $\mathbf{V}^{(-3)}$ beam*

The absolutely other situation is observed for the polarization singularities in the $U$-beam group. The picture in Fig. 4d represents also a complex combination of three lemons and three stars at the angle $\alpha_o = 8^o$. When further increasing the angle $\alpha_o$, three stars gather together at the point $x = 0$, $y' = 0$ forming a degenerated polarization singularity with a topological index $\nu = -3/2$ (Fig. 4e). However, the rest three lemons form something like an asymmetric vortex-cloud shifted along the $x$-axis in the vicinity of the central point. Such a vortex structure does not change while growing the crystal length. The growth of the topological charge $|l|$ of the initial beam entails a linear displacement of the lateral C points along the $x$-axis for both the $V$- and $U$-groups of the beams.

Before studying the influence of the handedness of a circular polarization and a sign of the vortex topological charge on the transverse shift of optical vortices, let us note that the beam fields presented by the expressions (23)-(26) are not the only ones. We can construct new groups of fields with the help of changing the variable of differentiation and integration from $\overline{u}$ to $\overline{v}$ in eqs (11), (15) and (12), (13). Then the equations (11) and (15) can be rewritten in the form:

$$V_+^{(-l),(o,e)} = \mp N_l^V \partial_v^{l-1} \partial_u \Psi_0^{(o,e)}$$
$$V_-^{(-l),(o,e)} = N_l^V \partial_v^{l-1} \partial_v \Psi_0^{(o,e)}, \qquad (29)$$

$$V_+^{(l),(o,e)} = \mp N_l^V \partial_v^{l-1} \partial_u \Psi_0^{(o,e)}, \; V_-^{(l),(o,e)} = N_l^V \partial_v^{l-1} \partial_v \Psi_0^{(o,e)}, \; (29)$$
$$\overline{U}_+^{(-l)} = N_l^U \partial_u^l (\overline{G}_+^{(o)} + \overline{G}_+^{(e)}), \; \overline{U}_-^{(-l)} = N_l^U \partial_u^l (\overline{G}_-^{(o)} + \overline{G}_-^{(e)}), (30)$$

where $\overline{G}_+^{(o,e)} = \pm\int_{\overline{v}}^{\infty}\partial_u \Psi_0^{(o,e)}\,dv$, $\overline{G}_-^{(o,e)} = -\int_{\overline{v}}^{\infty}\partial_v \Psi_0^{(o,e)}\,dv$

Besides, our requirement is now: the RHP component of the initial field vanishes at the z=0 plane: $\overline{V}_+(x,y,z=0) = 0$, $\overline{U}_+(x,y,z=0) = 0$. Then, the beam components get the form:

$$\overline{V}_+^{(l)} = -N_l^V \sum_{p=1}^{l}\binom{l}{p}\left(\frac{\alpha_o z_o}{w_0}\right)^{l-p}\partial_v^{p-1} \times$$
$$\times(\partial_u \Psi_0^{(o)} - \partial_u \Psi_0^{(e)}) + \left(\frac{\alpha_o z_o}{w_0}\right)^l \overline{U}_+^{(0)} \; ; \quad (l \neq 0) \; (31)$$

$$\overline{V}_-^{(l)} = \left[\frac{x + i(y - \alpha_o z)}{w_0 \sigma_o}\right]^l \Psi_0^{(o)} +$$
$$+ \left[\frac{x + i(y - \alpha_e z)}{w_0 \sigma_e}\right]^l \Psi_0^{(e)} \qquad (l \neq 0) \; (32)$$

$$\overline{U}_+^{(-l)} = N_l^U \sum_{p=0}^{l}\binom{l}{p}\left(-\frac{\alpha_o z_o}{w_0}\right)^{l-p}\partial_u^p(\overline{G}_+^{(o)} + \overline{G}_+^{(e)}), \; (33)$$

$$\overline{U}_-^{(-l)} = \left[\frac{x - i(y - \alpha_o z)}{w_0 \sigma_o}\right]^l \Psi_0^{(o)} + \left[\frac{x - i(y - \alpha_e z)}{w_0 \sigma_e}\right]^l \Psi_0^{(e)}.(34)$$

The LHP component of the $V$-group of the initial beams carries over positively charged vortices whiles the $U$-group – negatively charged ones. Comparing with the field (23)-(26) we can infer that the handedness of the circular polarization of the beams and the vortex topological charges are tightly bound with each other in the crystal-traveling beams. Computer simulation showed that the directions of the vortex transversal shift both in $V$- and $U$-groups *are exclusively defined by the handedness of the initial circular polarization*. Otherwise, the vortex transversal shift has the same direction for the $V$-beam group with the RHP and the $l < 0$ topological charge and the $U$-beam group with the RHP and the $l > 0$ topological charge (just as for the $\overline{V}$-beam group with the LHP and $l > 0$ and the $\overline{U}$-beam group with the LHP and $l < 0$).

Linearly polarized vortex-beams represent superpositions of the circularly polarized ones: $\mathbf{F}_X^{(-l)} = \mathbf{V}^{(-l)} + \overline{\mathbf{U}}^{(-l)}$ and $\mathbf{F}_Y^{(-l)} = -i(\overline{\mathbf{V}}^{(-l)} - \mathbf{U}^{(-l)})$ with a linear polarization of the initial beam directed along the $x$- and $y$-axis, respectively. Naturally, these vortex-beams are non-uniformly polarized near the beam core in the asymptotic case. Fig. 5 illustrates behavior of the polarization singularities inside the core of the vortex beam $\mathbf{F}_X^{(-3)}$ after splitting. At the inclination angle $\alpha_o = 7^o$, three lemons and three stars are grouped around the beam axis. As the angle $\alpha_o$ increases, two stars flow together forming a degenerated star at the beam axis $y' = 0$, $x = 0$ at the angle $\alpha_o = 10^o$ while the asymmetric cloud of three lemons are shifted to the left and one star is shifted to the right. Thus, a magnitude of the relative shift between the polarization singularities is doubled. In the RHP components, we will observe the cloud of optical vortices shifted to the left whereas the LHP component comprises the centered double-charged vortex and one singly-charged vortex shifted to the right.

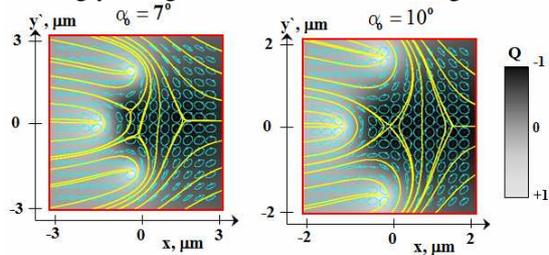

*Fig.5 (online) Polarization singularities inside the core of the linearly polarized vortex-beam $\mathbf{F}_x^{(-3)}$ (l=-3): z=2 cm, $w_0$=30 μm*

### III.3 The experiment

A series of nontrivial theoretical results described above needs an experimental basis. First of all, this relates to a different behavior of RHP and LHP components in the splintered beams that results in a non-uniformly polarized field distribution in the vicinity of the beam core. It is important to note that the polarization heterogeneity of the paraxial beams split by the uniaxial crystal is inherent to all singular beams irrespective of a magnitude of their topological charges. Although the area of the polarization heterogeneity increases when growing the vortex topological charge *l*, light intensity decreases very quickly in this area. Besides, a high-order optical vortex embedded in the beam is of an unstable structure that can be destroyed by a very small external perturbation [40]. Naturally, experimental measurements with off-axis high order vortex-beams transmitting through a series of boundary faces of the optical elements in the real experimental set-up are accompanied by a very high experimental error. As a result, we restricted ourselves to the experiments with singular beams bearing singly-charged optical vortices. In the article [13], we have experimentally considered the vortex-beam behavior in a uniaxial crystal in the vicinity of the indistinguishability boarder $\alpha_o = \alpha_{ind}$ (or $z = z_{ind}$)]. In the given section, we will focus our attention on the beam structure far from this border when the beams are separated and their mutual interference vanishes. We will focus our attention on the transformations of the shape of the C-lines in the vicinity of the beam core caused by changing the initial circular polarization and the vortex topological charge.

The sketch of the experimental set-up is shown in Fig.6. The non-singular beam from the Ne-Ne laser $(\lambda = 0.6328\mu m)$ is transformed into a vortex-beam with a topological charge $l = \pm 1$ by a computer generated hologram (Tr). The sign of the topological charge is defined by the direction of the centered fork in the diffractive grating and can be switched by a simple rotation of the transparent at the angle $180^o$. Diffraction orders after the computer generated hologram are clearing by the diaphragm (D). The polarizer (Pol) and the quarter-wave retarder $(\lambda/4)$ insert a circular polarization into the beam. The handedness of the circular polarization is defined by the direction of the axes of the $\lambda/4$ plate and can be converted into opposite one by a simple rotation of the $\lambda/4$ plate axes at the angle $90^o$. Further the beam is focused by a lens (L) with the focal length $f = 5cm$ into the LiNbO$_3$ crystal at the angle $\alpha_{in}$ to the crystal optical axis **C** (the crystal length is about z=2 cm). The crystal is positioned on a rotary table that enables us to rotate the crystal with the angle precision about $0.03^o$. The beam after the crystal is collimated by the diaphragm (D) and 20$^x$ microobjective (MO). After passing through a quarter-wave retarder and the polarizer, the beam is detected by the CCD camera. The optical elements positioned after the crystal are mounted on the special 3D-optical table that permits us to tune up the beam image at the CCD camera after rotating the crystal. We could measure the specific Stokes parameters at each pixel of the beam image at the CCD camera in accordance with a standard technique [7, 13]. The spatial resolution in the beam image was about $1.5\mu m$. The position of the origin on the x-y plane is defined as a center of gravity of the beam at the CCD camera plane for each angle $\alpha_{in}$ to within $1.7\mu m$ provided that the initial beam is a linearly polarized. The described above technique does not permit us to measure a magnitude of the asymptotic transverse shift of the vortices $\Delta x_V$ (see Sections II and IV). Nevertheless, we can study experimentally major features of the fine polarization structure of the beam core and bring to light major tendencies of the vortex-beam behavior in each circularly polarized component when tilting the beam. We measured the positions *x* and *y* of the C-points for each angle $\alpha_{in}$ within to $1.5\mu m$ on the base of standard method [7, 13]. We started to measure the C-point positions at the beam cross-section when the light intensity between the beams was 10 times as small as the intensity at the beam maximum. This corresponds to a rather well splitting of the partial beams. A typical map of the polarization distribution is shown in Fig.7. We can see here standard patterns of the star and the lemon around the C-points. When tilting the beam, the C-points start to rotate and a distance between them changes. Their positions trace a complex trajectories in the space: $x, y, \alpha_{in}$.

At first, we plotted trajectories traced by the lemon and the star inside the core of the ordinary singular beam with a right hand circular polarization and a negative topological charge $l = -1$ at the input crystal face shown in Fig.8 (the lower partial beam in the Fig.6). The star is associated with a vortex in the RHP component whereas the lemon corresponds to the vortex in the LHP component. They move along spiral-like trajectories. The radii of their rotation decrease gradually. Note that the handedness of the C-line rotations is the same. (However, the handedness of both trajectories changes its sign in the extraordinary beams propagating at the angle $\alpha_e$ (the upper partial beam in the Fig.6)).

The star is rotated around the $\alpha_o$ axis while the axis of rotation of the lemon is asymptotically approach to the axis of rotation of the star. After the angle $\alpha_{in} \approx 12^o$, the trajectories draw together at the distance lesser than $2\mu m$ and are experimentally perceived as one line. When changing a sign of the initial circular polarization to the opposite one, the star and the lemon in Fig. 7 trade places. The lemon is moved now along a spiral-like trajectory rotating around the $\alpha_o$ axis. The trajectory of the star is shifted to the positive direction of the x-axis approaching gradually to the trajectory of the lemon. The switching of a sign of the initial vortex to the opposite one does not change essentially the form of the C-lines. Thus, the direction of the transverse shift of the vortices is exceptionally defined by a handedness of the initial circular polarization. At the other hand, the direction of the transverse shift of the vortices changes to the opposite one when changing the sign of the inclination angle $\alpha_o$.

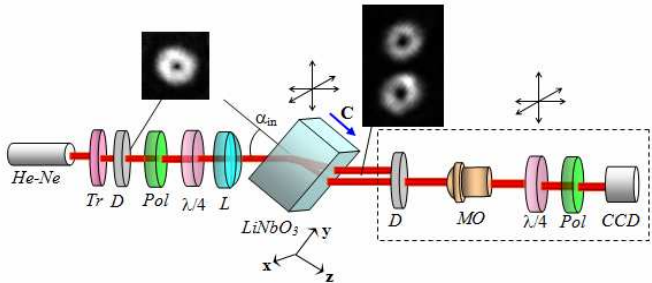

*Fig. 6. (online) The sketch of the experimental set-up: (He-Ne) –a laser, (Tr) – a computer-generated hologram, (Pol) – a polarizer, $\lambda/4$ - a quarter-wave retarder, (L) – a lens with f=5 cm, (LiNbO$_3$) – a crystal, (D)- a diaphragm, (MO) – a 20$^x$ microobjective; (CCD) –a CCD camera, C is a unit vector of the crystal optical axis*

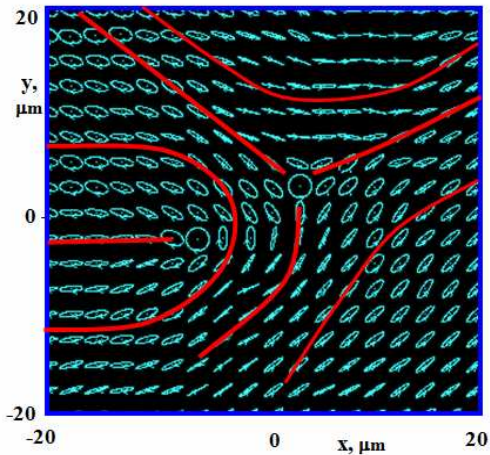

*Fig. 7(online) Typical map of the polarization distribution in the vicinity of the beam core with the initial RHP polarization and negatively charged vortex at the crystal input*

For comparison, we introduce theoretical trajectories in Fig.9 plotted for much the same parameters of the beam and the crystal. (The equation for the C-lines is derived from the requirement: the first and the second specific Stokes parameters vanish or $s_1(x,y,z,\alpha_o) = E_+ E_-^* + E_+^* E_- = 0$, $s_2(x,y,z,\alpha_o) = i(E_+ E_-^* - E_+^* E_-) = 0$). The theoretical C-lines for the $\mathbf{V}^{(-)}$-beam in Fig.9 also have all major features of the experimentally observed trajectories and are in a good qualitative agreement with the experimental curves in Fig.8.

The behavior of the lemon and the star in the initially x-linearly polarized beam $\mathbf{F}_x$ is shown in Fig.10. The C-lines have also a spiral-like form. However, both trajectories are symmetrically shifted along the positive and negative directions of the x-axis. In contrast to the circularly polarized $\mathbf{V}^{(-)}$-beam, the tilted linearly polarized $\mathbf{F}_x$ beam has a vanishingly small intensity of the orthogonal component. As a result, the interference between the ordinary and extraordinary beams is experimentally observed only for very small inclination angles. The C-lines of the linearly polarized $\mathbf{F}_x$ beam oscillates much slower and approach to each other very quickly. Nevertheless, we observe distinctly two trajectories drawing together with relatively slow oscillations. After the angle $\alpha_{in} \approx 7^o$, the trajectories are experimentally undistinguished and we cannot judge about their asymptotic behavior. Comparison of the curves in Fig.9 for the $\mathbf{F}_x$ and Fig.10 shows their good qualitative agreement.

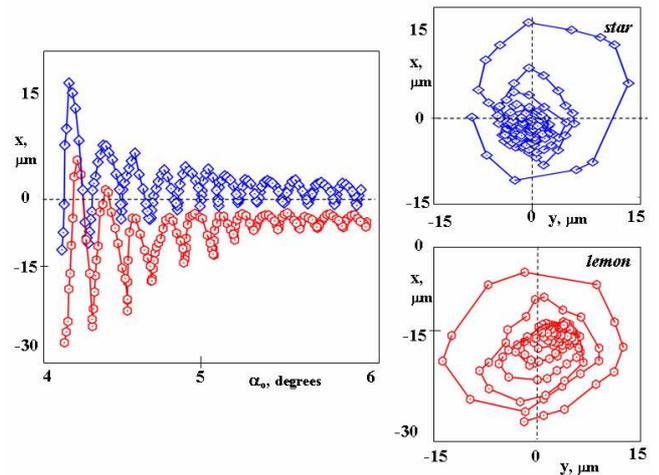

*Fig. 8 (online) The fragment of the C-lines in the vortex-beam at the z=0, transmitting at the angle $\alpha_o$ for the RHP initial beam with a topological charge l=-1, $w_0 \approx 50\mu m$, $z \approx 2cm$*

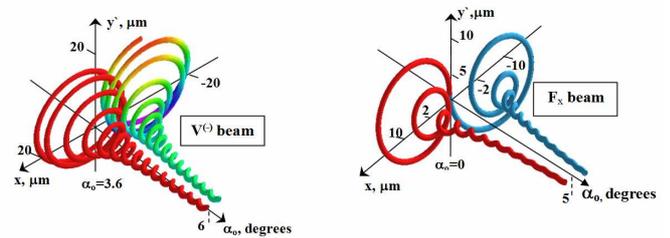

*Fig. 9 (online) A computer simulation of the C-line behavior in the $\mathbf{V}^{(-)}$ and $\mathbf{L}_x$ beams*

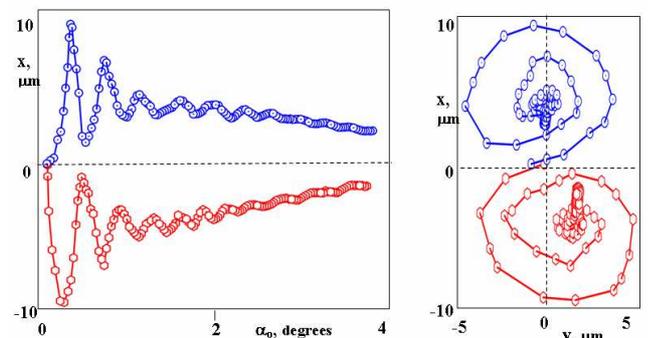

*Fig. 10 (online) The C-lines in the vortex beam with the initial linear polarization: $l = -1$  $w_0 \approx 50\mu m$, $z \approx 2cm$*

The vortex transverse shift in the singular beams stimulates inevitably the transverse shift of the circularly polarized beam components and distortion of their cross-sections.

# IV The transverse shift, angular momentum and deformation of the beam cross-section

## *4.1 The angular momentum*

As we have shown in Section II the transverse shift of the vortex-beam depends on 1) the beam depolarization and 2) transformation of the beam cross-section caused by dislocation reactions. The depolarization processes in the vortex-beam can be considered on the base of the Stokes parameters:

$$S_0 = \Im = \Im_+ + \Im_-, \quad S_1 = \int_{-\infty}^{\infty} dx \int_{-\infty}^{\infty} dy \left( E_+ E_-^* + E_+^* E_- \right),$$

$$S_2 = i \int_{-\infty}^{\infty} dx \int_{-\infty}^{\infty} dy \left( E_+ E_-^* - E_+^* E_- \right), \quad (35)$$

$$S_3 = \int_{-\infty}^{\infty} dx \int_{-\infty}^{\infty} dy \left( E_+ E_+^* - E_-^* E_- \right) = \Im_+ - \Im_-, \quad (36)$$

here the symbol $(^*)$ stands for a complex conjugation. $\pm \Im$ are intensities of RHP and LHP components, respectively. The magnitude

$$S_z = \frac{S_3}{\Im} = \frac{\Im_+ - \Im_-}{\Im} = I_+ - I_- \quad (37)$$

describes the spin angular momentum of the vortex-beam. The polarization degree can be presented as

$$P = \frac{\sqrt{S_1^2 + S_2^2 + S_3^2}}{\Im}. \quad (38)$$

Fig.11 demonstrates the polarization degree $P(z,l)$ as a function of the crystal length z for the beams with topological charges $|l| = 0, 1, 2, 3$. First of all, the curves $P(z,l)$ coincide with each other for all beam groups ($V, U, \overline{V}, \overline{U}$) with the same topological charges $|l|$. The polarization degree $P$ decreases monotonically only for the vortex-free initial beam. The rest curves $(l \neq 0)$ have an oscillation character. A number of oscillations increases with growing the topological charge $|l|$. This comes out from the fact that dislocation reactions in tilted beams progress more intensively for the vortex beams with a large topological charge. The magnitude $P$ vanishes when increasing the crystal length.

The SAM of the vortex beams $S_z$ is tightly bound with the polarization degree $P$ and also vanishes at the relatively large length z. The curves shown in Fig.12 demonstrates behavior of the SAM for the $\mathbf{U}^{(0)}$ and $\mathbf{V}^{(-3)}$ beams. The oscillations of the SAM in the $\mathbf{V}^{(-3)}$ beam die down periodically before vanishing while the oscillations of the vortex-free $\mathbf{U}^{(0)}$ beam decrease monotonically. In both cases the oscillations have envelopes in the form of the polarization degree $P$.

The OAM of the paraxial beam $L_z$ is calculated by mean of the expression:

$$L_z = -\frac{i}{\Im} \int_{-\infty}^{\infty} dx \int_{-\infty}^{\infty} dy \ \mathbf{E}^* \left( x \partial_y \mathbf{E} - y \partial_x \mathbf{E} \right), \quad (39)$$

It can be presented as a sum of the OAM of the RHP and LHP beam components $L_z = L_z^+ + L_z^-$. The curves shown in Fig. 13 describe the OAM evolution for the $\mathbf{U}^{(0)}$, $\mathbf{U}^{(3)}$ and $\mathbf{V}^{(-3)}$ beams The $\mathbf{U}^{(3)}$ and $\mathbf{V}^{(-3)}$ beams have opposite signs of the topological charges $l = +3$ and $l = -3$, respectively, and the same handedness of the circular polarization $\sigma = +1$ at the initial plane z=0. All the curves $L_z(\alpha_o)$ oscillate synchronically with the curves $S_z(\alpha_o)$ (see Fig. 11).

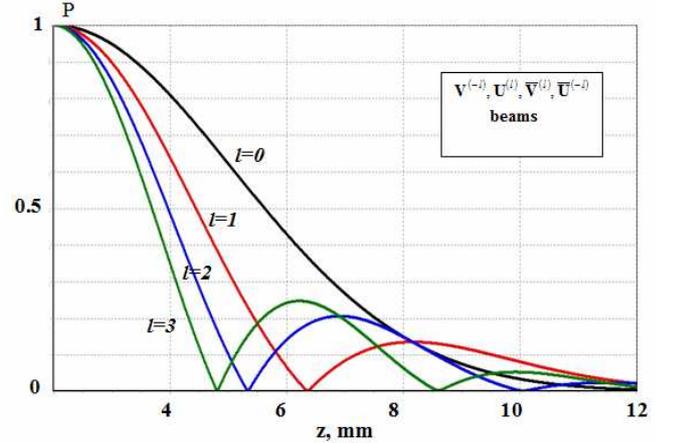

*Fig. 11 (online) Polarization degree P as a function the length z of the LiNbO$_3$ crystal for the $\mathbf{V}^{(-l)}, \mathbf{U}^{(l)}, \overline{\mathbf{V}}^{(l)}, \overline{\mathbf{U}}^{(-l)}$ beams with w$_0$=50 μm and the inclination angle $\alpha_o = 10^o$*

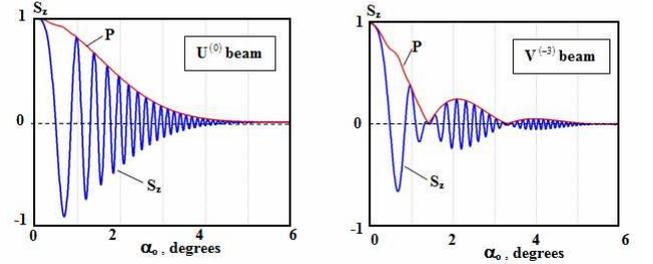

*Fig. 12 (online) The SAM S$_z$ and the polarization degree P of the vortex-free beam $\mathbf{U}^{(0)}$ and the vortex-beam $\mathbf{V}^{(-3)}$ as a function of the inclination angle $\alpha_o$ at the z=2 cm crystal length*

However, the sections of the curves in Fig. 11 where the SAM increases, corresponds to the section of the curves in Fig. 13 where the OAM decreases and vice versa. Such behavior of the curves is a result of the spin-orbit coupling in the uniaxial crystal. The oscillations fade away gradually near some asymptotic value $L_z^{(asymp)}$ of the OAM unequal to the initial OAM of the beam. The curves $L_z(\alpha_o)$ for the $\mathbf{U}^{(3)}$ and $\mathbf{V}^{(-3)}$ beams can be reconciled by the displacement of one of the curves along the ordinate axis at the distance $L_z = 6$. Note that the vortex-free $(l = 0)$ initial beam $\mathbf{U}^{(0)}$ with a zero OAM $L_z(z = 0) = 0$ acquires also the asymptotic orbital angular momentum $L_z^{(asymp)} = +1$. However, the total AM flux is conserved and depends on neither the crystal length z nor the angle $\alpha_o$:

$$l + \sigma = L_z(z, \alpha_o) + S_z(z, \alpha_o) = L_z^{(asymp)}(z \to \infty), \quad (40)$$

where $\sigma = \pm 1$ is a handedness of the circular polarization of the initial beam. At the large crystal length, the beam is depolarized. It's the SAM vanishes: $S_z(z \to \infty) = 0$ while the OAM gets an additional magnitude. Such an additional AOM originates from an asymptotic transverse shift $\Delta x_T$ of the beam.

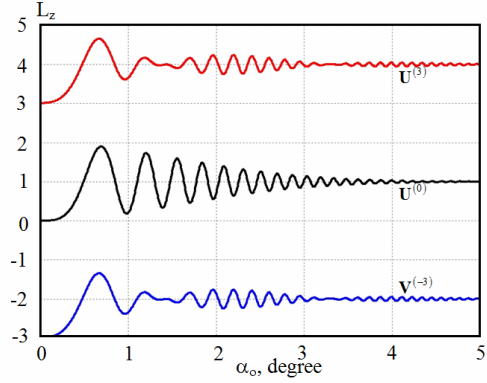

*Fig.13 (online) Transformations of the OAM $L_z$ for the beam fields $U^{(3)}$, $U^{(0)}$ and $V^{(-3)}$; $w_0=50\ \mu m$, $z=2\ cm$*

### 4.2 The transverse shift

We calculated the shift of the center of gravity of the beam on the base of the standard expressions:

$$x_C = \frac{1}{\Im}\int_{-\infty}^{\infty}dx\int_{-\infty}^{\infty}dy\ x\ |\mathbf{E}(x,y,z,\alpha_o)|^2,$$

$$y_C = \frac{1}{\Im}\int_{-\infty}^{\infty}dx\int_{-\infty}^{\infty}dy\ y\ |\mathbf{E}(x,y,z,\alpha_o)|^2. \quad (41)$$

The chains of the events of the vortex birth and annihilation caused by dislocation reactions in the tilted vortex-beams force the center of gravity trace intricate space trajectories for each of the circularly polarized components. The typical trajectories are shown in Fig. 14 for the $\mathbf{V}^{(-3)}$ vortex-beam. The amplitude of vibrations of the trajectory depends on the crystal length $z$ and the inclination angle $\alpha_o$. The amplitude has large value inside the angle range from $\alpha_o = 0$ to the indistinguishability border $\alpha_o = \alpha_{ind}$. In the vicinity of the value $\alpha_o = \alpha_{ind}$, the vibrations die down because a part of vortices that take place in the reconstruction of the beam core, leaves the area of dislocation reactions. Then the vibrations are resumed again but with essentially smaller amplitude while their frequency increase very much. Finally, the vibrations fade away at the relatively large angles $\alpha_o$ (or the crystal length $z$). However, the positions of the center of gravity of the $V_+^{(-3)}$ and $V_-^{(-3)}$ components are shifted along the x-axis at the distance: $\Delta x^- = -2/(k\alpha_o)$ while the shift on the orthogonal plane vanishes $\Delta y^- = 0$. At the same time, the center of gravity of the total beam $\mathbf{V}^{(-3)}$ is shifted only at a half of this distance: $\Delta x_T = \Delta x^-/2$ but for all that the $\bar{V}_+^{(-3)}$ component is not shifted.

The circularly polarized components of the $\mathbf{F}_X^{(-2)}$ beam with the initial linear polarization directed along the *x*-axis and the vortex topological charge *l*=-2 are shifted in opposite directions (see Fig. 15) so that in the asymptotic case the transverse shift between circularly polarized components is doubled $\Delta x_C^{(V)} = 4/(k\alpha_o)$.

It is noteworthy to remark that the magnitudes of the asymptotic transverse shifts $\Delta x_T$ of the center of gravity for the circularly polarized component of the $V, U, \overline{V}, \overline{U}$ beams are the same for different magnitudes of the initial vortex topological charges *l* although the transverse shift of the vortices in these components is proportional to the vortex topological charge *l*.

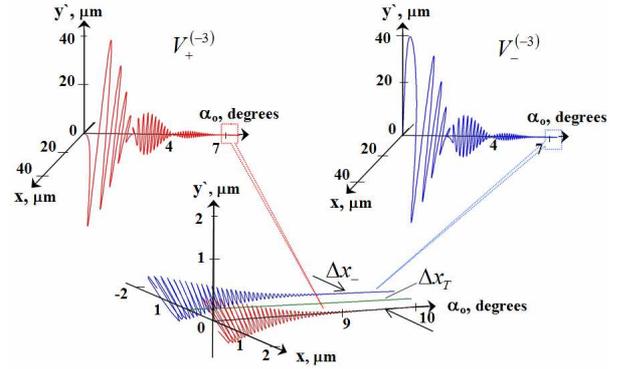

*Fig. 14 (online) The trajectories of the center of gravity of the circularly polarized components of the $\mathbf{V}^{(-3)}$ beam; $z=2\ cm$, $w_0 = 50\ \mu m$*

Indeed, the equation (40) in common with the expression (2) for the angular momentum flux show that the sum of the SAM and OAM at the initial plane z=0 is equal to the asymptotic AM flux at the $z\to\infty$, i.e $l+\sigma = \frac{1}{2}l_+ + \frac{1}{2}l_- - k\alpha_o\Delta x_T$. However, the vortex topological charges in the RHP and LHP components equal each other $l_+ = l_- = l$ in the asymptotic case so that

$$k\alpha_o\Delta x_T = -\sigma. \quad (42)$$

This magnitude is a constant of the uniaxial crystal. Note also that the transverse shift derived from the conservation low for the total linearly polarized $\mathbf{F}_{X,Y}^{(l)}$ beams, equals zero although it has the entirely definite magnitude in each circularly polarized component. The transverse shift in the linearly polarized components vanishes too.

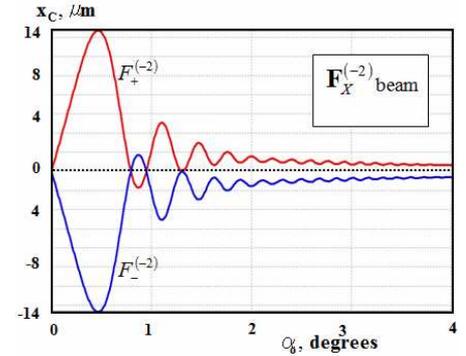

*Fig. 15 (online) Transverse shift $x_C$ of the circularly polarized components of the linearly polarized $\mathbf{F}_X^{(-2)}$ vortex-beam with $l = -2$, $w_0=50\ \mu m$, $z=2\ cm$*

### 4.3 The deformation of the beam cross-section

The asymmetric vortex destruction and recovery in the paraxial beam considered in Sec. III causes not only the transverse shift but also distortion of a circular symmetry of the beam cross-section. Generally speaking, a uniaxial crystal deforms the initially circular cross-section of the paraxial extraordinary beam when propagating perpendicular to the crystal optical axis [37] even without taking into account the vortex structure of the beam. Complex behavior of such deformation in tilted paraxial beams was remarked in the article [3]. A circularly polarized beam propagating along the crystal optical axis does not experience an elliptical deformation. At the same time, a linearly polarized Laguerre-Gaussian beam transmitting along the crystal optical axis undergoes a relatively strong deformation, its magnitude increasing as the beam propagates along the crystal [38]. Contribution of the transverse shift to the beam deformation is of the object of a special investigation. However, in the given Section we will consider only some features of such a complex process.

The magnitude of the cross-section deformation can be estimated by means of the *mean square width* of the paraxial beam (see, e.g., [37]):

$$W_\pm^2(\varphi,z,\alpha_o,l) = \frac{1}{\mathfrak{S}_\pm} \int_{-\infty}^{\infty} dx \int_{-\infty}^{\infty} dy' (x\cos\varphi + y'\sin\varphi)^2 \left|E_\pm^{(l)}\right|^2, \quad (43)$$

where $\varphi$ is the azimuthal angle in the referent frame $x0y'$. Later on we will assume that the ordinary and extraordinary beams are separated in the asymptotic case and take into account only the field of the ordinary beam both in RHP and LHP beam components. The curve in Fig. 16 illustrates typical behavior of the mean square width $W_\pm = \sqrt{W_\pm^2}$ as a function of the angle $\varphi$. Variations of the magnitude $W_+$ of the RHP component of the $U^{(3)}$ beam vanishes in the asymptotic case while the magnitude $W_-$ changes periodically with the angle $\varphi$ between the extremal value $W_{max}$ and $W_{min}$. The same behavior is observed for the $V^{(-3)}$ beams. To estimate a relative deformation of the beam we introduce a subsidiary parameter in the form: $\overline{W}_\pm = \frac{1}{\pi}\int_0^\pi \sqrt{W_\pm^2} d\varphi$ so that the deformation $D$ relative to the asymptotic transverse shift $\Delta x_T$ is

$$D(z,\alpha_o,l) = \frac{W_{max} - \overline{W}}{\Delta x_T}. \quad (44)$$

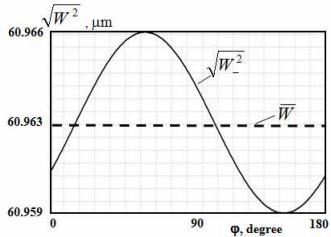

Fig. 16 The mean square width $\sqrt{W_-^2}$ of the LHP component of the $U^{(3)}$ beam: as a function of the angle $\varphi$: $\alpha_o = 10^o$, $z = 2\,cm$, $w_0 = 50\,\mu m$

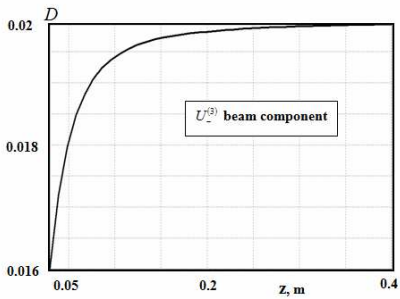

Fig. 17 Deformation D of the LHP component $U_-^{(3)}$ as a function of the crystal length z: $\alpha_o = 10^o$, $w_0 = 50\,\mu m$

The curve $D(z)$ shown in Fig. 17 illustrates deformation of the cross-section of the $\overline{U}^{(3)}$ beam component along the crystal length z. The beam deformation changes very slowly along the crystal and its magnitude is only $D = 0.02$ at the crystal length $z = 0.4\,m$. Absolutely other situation is observed for the linearly polarized *F*-beams. The cross-sections of the linearly polarized beams are distorted even for the on-axis beams $\alpha_o = 0$ [38] when the transverse shift vanishes. Non-monotonous variations of the deformation parameter $D_{max} - \overline{D}$ of the $\mathbf{F}^{(-3)}$ beam with the initial *x*-linear polarization and a triple-charged (*l*=-3) optical vortex shown in Fig. 18 are connected with a complex structure of the RHP and LHP components of the beam transmitting along the crystal optical axis. Figures 18a-18d demonstrate the intensity distributions of the LHP component at different crystal lengths. At the relatively small length (about 2 cm), the beam deformation is vanishingly small (Fig.18a). As the beam propagates along the crystal, a fine asymmetric vortex structure becomes faintly visible (Fig. 18b) on the background of a uniform intensity distribution. The distortion amounts to its maximum value at the distance z=0.25 m (Fig. 18c). Then the asymmetric vortex distribution is partially smoothed by appearing new outlying vortices and the beam distortion decreases (Fig. 18d). The inclination of the linearly polarized beam changes the picture as a whole. Dislocation reactions sort out peripheral vortices so that after transferring through the indistinguishability border $\alpha_o = \alpha_{ind}$ the optical vortices gather together in the vicinity of the beam core (see Fig. 3). When tilting the beam the major role in the beam deformation shown in Fig. 19 starts to play the transverse shift. Although the asymptotic transverse shift $\Delta x_T$ does not depend on the crystal length, the beam deformation induced by them increases along the crystal length at the expense of the diffraction process. Differences in behavior of the curves for RHP and LHP components at the relatively small crystal length in Fig. 19 are caused by a different distribution of the optical vortices (see Fig. 3) in the cross-section of these components.

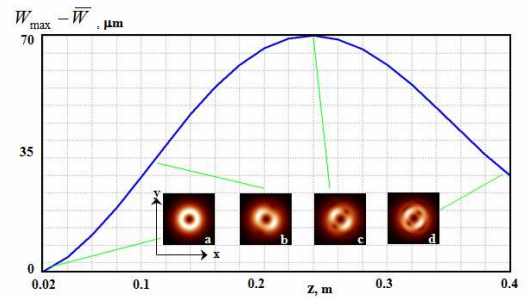

Fig. 18 (online) Transformation of the parameter $W_{max} - \overline{W}$ along the crystal length z for the LHP component of the linearly polarized on-axis $\mathbf{F}_X^{(-3)}$ beam: $\alpha_o = 0$, $w_0 = 50\,\mu m$ and intensity distributions (a)-(d) for its RHP component at the different crystal lengths

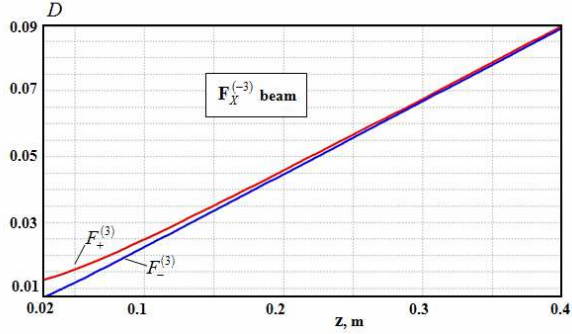

*Fig. 19.(online) Deformation D of the RHP and LHP components of the $\mathbf{F}_X^{(-3)}$ beam along the crystal length z:*
$\alpha = 10^o$, $w_0 = 50\,\mu m$

## V. Conclusions

We have discussed the destruction and recovery of high order vortex-beams in the unbounded medium of a uniaxial crystal when the beam propagates at a small angle to the crystal optical axis. We paid a special attention to the transverse shift of the beam induced by the homogeneous anisotropic medium and the beam deformation caused by the transverse shift in one of the circularly polarized components. It has been shown that an inclination of the beam relative to the crystal optical axis is tightly connected with a global reconstruction of the vortex structure. For example, a RHP singular beam bearing the optical vortex with a topological charge equal to *-l* at the crystal input stimulates appearance of the LHP singular beam with a topological charge *–l+2* when propagating along the crystal optical axis. When tilting the beam the LHP component of the beam loses two positively charged optical vortices while the RHP component keeps its former vortex structure. At the first glance it seems that both circularly polarized components carries over now identical optical vortices. However, we have shown that a fine structure of the beam core in the RHP and LHP components is different. All optical vortices gather together at the axis of the RHP component forming the *l*-charged optical vortex. At the same time, only *l-1* vortices gather together at the axis of the LHP component while one singly charged vortex is shifted at the distance $\Delta x_V = -2l\sigma/(k_o\alpha_o)$ along the direction perpendicular to the inclination plane of the beam. Such a vortex shift stimulates inevitably the transverse shift of the vortex-beam. This transverse shift of the beam is a result of the conservation law of the total angular momentum flux along the crystal optical axis. Indeed, when propagating the beam its polarization degree vanishes and, consequently, the spin orbital momentum vanishes too. At the same time, the loss of two vortices entails the transformation of the orbital angular momentum. To conserve the total angular momentum, the beam experiences a transverse shift equal to $\Delta x_T = -\sigma/(k_o\alpha_o)$ at the asymptotic case when crystal length is very large. Our detailed analysis showed that the RHP component is not subjected to the transverse shift $\Delta x^+ = 0$ at a large crystal length $z \to \infty$ whereas the LHP component gets a doubled transverse shift $\Delta x^- = 2\Delta x_T$ for the RHP initial beam. Such a transverse shift does not depend on a sign and magnitude of the vortex topological charge *l* in contrast to the transverse shift of the refracted and reflected singular beams at the boundary face of two homogeneous media [27, 28]. However, it changes its direction to the opposite one when switching the handedness of the initial circular polarization and changing a sign of the inclination angle $\alpha_o$. In the initially linearly polarized beam, both circularly polarized components experience the transverse shift in opposite directions. In fact, we deal with *the beam quadrefringence* in a uniaxial, homogeneous anisotropic medium. The first two beams is a result of the splitting of the total tilted beam into the ordinary and extraordinary ones. The appearance of the second two beams is caused by the transverse shift of one of the circularly polarized components in the initially circularly polarized vortex-beam or both circularly polarized components in the initially linearly polarized beam.

We have analyzed the deformation of the beam cross-section caused by the transverse shift. We revealed that deviations of the mean square width of the beam cross-section are vanishingly small for the RHP component whereas the LHP component experiences a deformation of the cross-section in the initially RHP beam. A relatively strong deformation experiences an initially linearly polarized vortex-beam. Although the deformation of the cross-section is a very small even relative to the transverse shift at real crystal lengths (1-2 cm), this increases very quickly when increasing the crystal length. Note that the asymptotic transverse shift does not depend on the crystal length. However, the diffraction process enhances the deformation caused by the transverse vortex shift.

### Acknowledgments
We are indebted to K. Yu. Bliokh for his valuable comments. We also thank E. Abramochkin for the useful discussion on the theoretical aspects of the work and B. Sokolenko for his help in the experiment.